\documentstyle[12pt,aaspp4]{article}
\def\day{{}$^{\rm d}$\llap{.}}

\begin{document}

\title{CCD OBSERVATIONS OF THE RR LYRAE VARIABLES IN THE GLOBULAR
CLUSTER NGC 5897}

\author{Christine M. Clement }

\affil{Department of Astronomy, University of Toronto \\
Toronto, ON, M5S 3H8, CANADA\\
electronic mail: cclement@astro.utoronto.ca}

\author{Jason F. Rowe }

\affil{Department of Physics and Astronomy, 2219 Main Mall\\ 
University of British Columbia,
Vancouver, BC, V6T 1Z4, CANADA\\
electronic mail: rowe@astro.ubc.ca}

\begin{abstract}
CCD observations for 10 (6 previously known and 4 newly discovered) of the
11 RR Lyrae variables in NGC 5897 have been analysed. 
The period-luminosity and period-amplitude plots indicate that the population
of RR Lyrae variables in NGC 5897 includes 3 fundamental mode, 4 
first-overtone and 4 second-overtone variables with mean periods
0\day 828, 0\day 459 and 0\day 343 respectively. The variables
have properties that are similar to some of the Oosterhoff
type II variables in $\omega$ Centauri.  Two of the new variables (V11 and V13)
were previously considered to be possible non-variables that lie within or
near the instability strip on the horizontal branch; both have $V$
amplitudes less than $0.2$ mag. There is a chance that one or two of 
the variables 
(V10 and possibly V3) might be anomalous Cepheids and not RR Lyrae stars.

\end{abstract} 
%
%

\section{INTRODUCTION}
NGC 5897 (C1514-208) is a moderately metal poor globular cluster with
[Fe/H] = $-1.80$ (Harris 1996). 
Color-magnitude diagrams for NGC 5897 have been
published by Sandage \& Katem (1968, hereafter SK), Ferraro et al.
(1992), Sarajedini (1992) and Testa et al. (2001) and all of them
show that the horizontal branch is predominantly blue. The cluster has
nine known variable stars: seven RR Lyrae (Wehlau 1990, hereafter 
W90), one red variable (Eggen 1972)
and one SX Phe (Wehlau et al.  1996).
W90 has shown that the RR Lyrae variables have a
very unusual period distribution; 
three have periods between 0\day 797 and 0\day 856, two at
0\day 420 and 0\day 454 and two more at 0\day 342 and 0\day 349.
As a result, the mean period of the three RRab 
variables is 0\day 828. 
The fact that the periods of the RRab variables in NGC 5897 are 
longer than 0\day 60 is not unusual  because
Clement et al. (2001) have shown that
for all clusters more metal poor than $-1.70$, the mean period of the 
RRab stars is greater than 0\day 60. 
What is unusual, however, is that $<P_{ab}>$ is greater than 0\day 80; the
next longest
$<P_{ab}>$ for a metal poor cluster is 0\day 708 for NGC 4833 which has
[Fe/H]$=-1.79.$\footnote{Generally, the RR Lyrae variables
in metal poor clusters 
have longer periods than the ones in metal rich clusters. 
However,
Pritzl et al. (2000) have recently shown that the two metal rich clusters
NGC 6388 and NGC 6441 have $<P_{ab}>$ of 0\day 71 and 0\day 76, respectively,
for their RR Lyrae variables. The variables in these two clusters seem to be
brighter than solar neighbourhood field RR Lyrae stars of comparable [Fe/H],
but the reason for this is not yet fully understood.}

Another curious feature of NGC 5897 is the fact that there appear to be
nonvariable horizontal branch stars located in its RR Lyrae gap. 
Both SK and W90 have shown that the stars
SK 116 and SK 174 have colors and magnitudes comparable to the RRc stars  
and that SK 120 lies in the instability
strip well to the red of the RRc stars and to the blue of the RRab stars.
Smith (1985) made a spectroscopic study of SK 120 and concluded
that the star is probably a cluster member. 
We therefore decided it would be worthwhile to observe
NGC 5897 with a CCD detector. 
It may turn out that some of these `apparent'
nonvariable stars are in fact varying, but with an amplitude
lower than can be readily detected in photographic studies.
All of the published photometric studies of the RR Lyrae variables in NGC 5897,
with the exception of a set of CCD observations of V2 by Wehlau et al. (1996),
are based on photographic data.

\begin{figure}
\epsscale{1.0}
\plotone{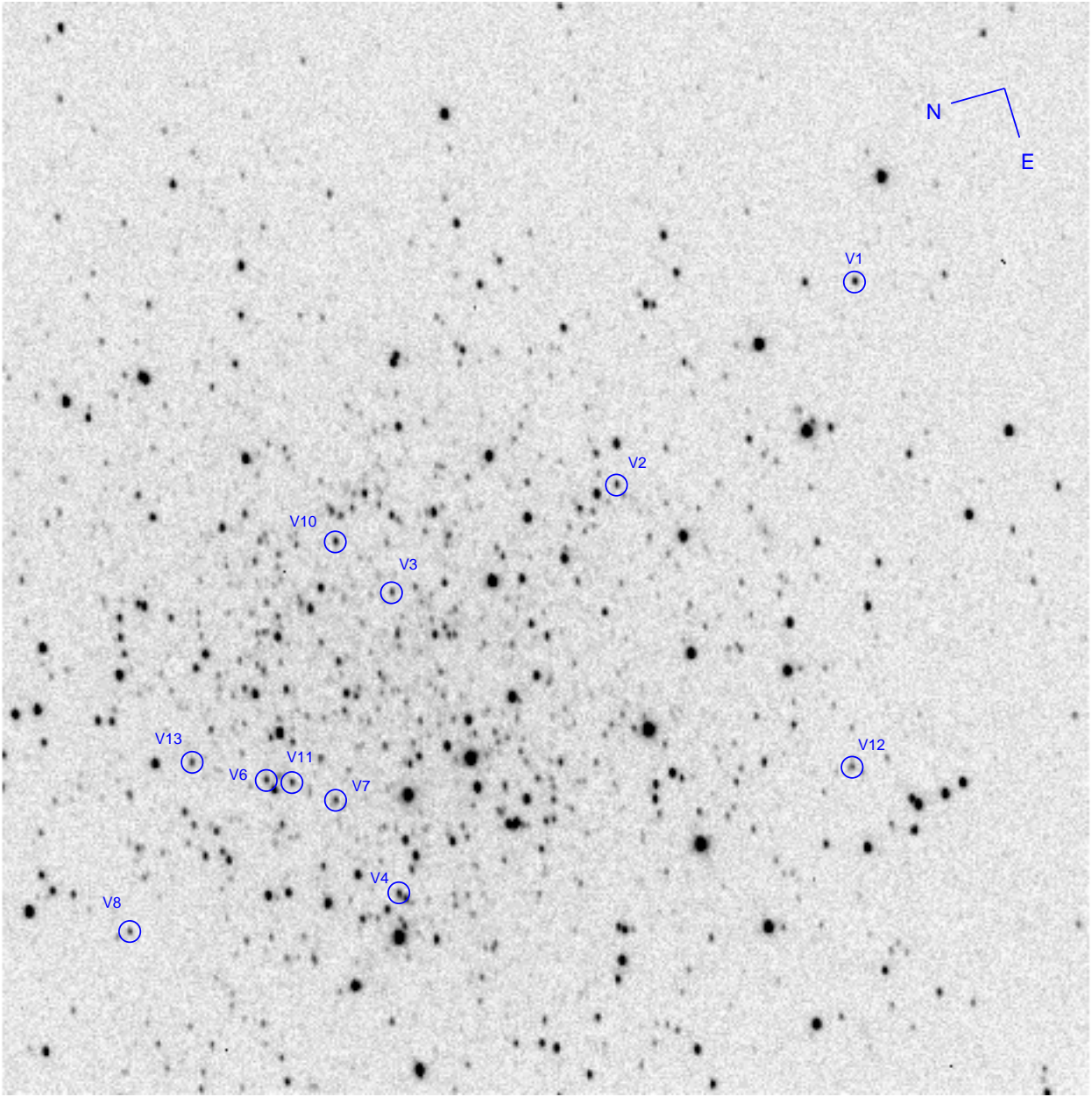}
\caption[]
{Identification chart for the RR Lyrae
variables in NGC 5897. The field of view is
$6.9 \times 6.9$ arcminutes. V1-4, 6-8 and 10-13 are labelled. West is at
the top and north on the left.
\label {Fig. 1}}
\end{figure}

%

\section{THE OBSERVATIONS}

Our investigation is based on 526 CCD frames obtained with the University of
Toronto's 61 cm ($f/15$) Helen Sawyer Hogg (HSH) telescope at the Las Campanas
Observatory of the Carnegie Institution of Washington. The observations were
made on six consecutive
nights (May 1 to 6, 1997) with
the Kodak KAF-4200 CCD system that Horch \it et al. \rm (1997) and
Slawson \it et al. \rm (1999) used with the same telescope in February
1997. The chip has 9 $\mu$m 
square pixels arranged in a 2033$\times$2048 format, but before readout,
the data were binned $2\times 2$ so that the effective pixel size for our 
data was 18 $\mu$m square.  
The field of view (shown in Figure 1) is approximately $6.9\times 6.9$ 
arcminutes and includes all of the RR Lyrae variables
studied by W90.
The frames were obtained through a V filter and all of the exposure times
were 2 minutes.  These short 
exposures were necessary because of severe tracking problems with the telescope.

The frames were cleaned using IRAF.
Then, for the photometric reductions, we combined the frames in groups of
three, so that we could obtain
photometry with a precision of $0.02$ mag  (or better) at V$\sim 16.5$, the 
level of the horizontal branch. 
In Table 1, we list a sample of the data obtained on the night of May 1
to illustrate how the frames were combined.  The photometric reductions were 
carried
out using the IRAF versions of 
DAOPhot-II and ALLSTAR-II 
and the stand-alone version of Stetson's (1987, 1993, 1994) 
ALLFRAME code. Approximately 1000 stars were identified on each 
frame.

In order to convert our instrumental magnitudes to standard V magnitudes,
we used the photometry of SK. Since none of
their photoelectric standards were in the field shown in Figure 1, 
we obtained 7 frames on the night of May 4 with the telescope offset to a 
position $5.7$ arcminutes further north. With this offset, we were able 
to include the SK photoelectric standard stars A, C, E,
K, M, N, O and P in the field of view. We determined the instrumental
magnitudes for the stars that were in the area overlapping the
two fields on the same
magnitude scale as the `offset' frames and then calculated magnitudes on the
same scale for all of the program stars. Finally, the transformation from
instrumental to standard V magnitudes was derived for SK's photoelectic
standards and then applied to all of the other stars.

%
%

\section{THE VARIABLES}
\subsection{Search for Variables and Period Determination}
   
The known RR Lyrae variables were identified and found to have a mean $V$
magnitude of $\sim 16.25$. To search for additional RR Lyrae variables,
we examined the observations for all of the stars with mean $V$ magnitude
between $15.75$ and $16.75$ (approximately 150 stars). 
We plotted light curves to check for periodic variations. As a result,
four new RR Lyrae variables were identified. We have numbered them
V10 to V13 and they are labelled in Figure 1. This increases
the total number of known RR Lyrae variables in NGC 5897 from seven to eleven.
In order to determine the periods for the 
new variables, we used Stellingwerf's (1978)
phase dispersion minimization technique to estimate an approximate
value and then
made small adjustments so that the observations for the different nights
would be well aligned. Our adopted periods are  listed in Table 2.
For the known variables, we adopted the periods listed by W90.
Also included in the table are  the SK number, the x,y position on the
frame in pixels,
the x,y position relative to the cluster center in arcseconds on the
system of Sawyer Hogg (1973), the mean $V$ magnitude, the $V$ amplitude along
with the standard deviation of the fit to equation (1)
and the Fourier phase differences $\phi_{21}$, $\phi_{31}$ along with their
standard errors.
The mean magnitudes and amplitudes were derived from a fit to a Fourier series
of the form:
\begin{eqnarray}
mag = A_0+\sum _{j=1,n} A_j \cos (j\omega t + \phi _j)  
                     \hskip 2mm 
\end{eqnarray}
where $\omega$ is ($2\pi$/period). The Fourier phase differences
($\phi_{ji}$) are ($\phi_j-j\phi_i$) from equation (1).
For each star, the epoch was taken as HJD $2,450,500$ so that $t$ in the 
equation refers to (HJD$-2,450,500$) and HJD represents the mean
heliocentric Julian date of the observation (in this case, the combination
of three frames).
Light curves, arranged in order of increasing period, are shown in Figure 2.
A sample of the observations for V1 is shown in Table 3. All of the 
observations for V1-3, V6-8 and V10-13 are presented in the electronic version
of the table.
V4 was not included in our study because its image was blended with that of
another star and as a result we were unable to obtain photometry with the
desired precision. In fact, many of the frames could not be used to study
the other variables either, because of the tracking problems.

\begin{figure}
\epsscale{0.6}
\plotone{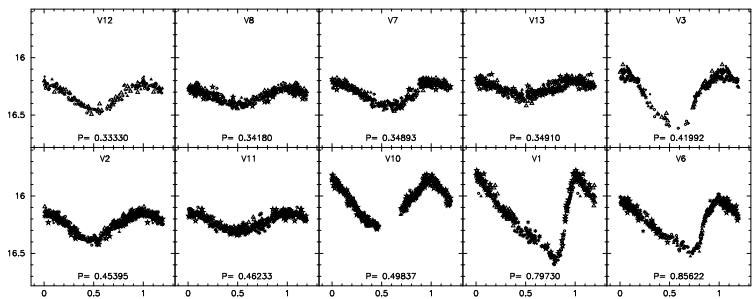}
\caption[]
{$V$ light curves for the 10 stars in our sample. The curves are arranged
in order of increasing period.
\label {Fig. 2}}
\end{figure}

It turns out that two of the new variables, V11 and V13 are
SK 120 and SK 116 respectively, 
apparent nonvariable stars that lie in or near the  RR Lyrae gap according to
SK and W90. Since the
$V$ amplitudes for these two stars are less than $0.20$ mag, 
they were not detected in photographic studies of the cluster. 
We also examined the magnitudes for SK 174, a star that W90 found to be 
near the blue edge of the instability strip,
but no systematic variation
was detected.  Its mean $V$ magnitude
is $16.392$ with a standard deviation of $0.029$. In comparison, the standard 
deviation of the previously
known variables ranged from $0.053$ for V8 to $0.215$ for V1.

The other two new variables, V10 and V12, have larger
amplitudes, but their periods are close to simple fractions of a day. This
probably accounts for the fact that they were not discovered previously. 
In fact, even in the present study, the phase coverage for V10 (P$\sim$0\day 5)
is incomplete. We considered the
possibility that V10 has a period twice as long, but concluded this to be
unlikely. The shape of its light curve is similar to that of three $\omega$
Centauri variables with similar periods, V47, V68
and V123 (OGLE ID 185, 73 and 169), 
observed by Kaluzny et al. (1997, hereafter K97). All of these
stars have an inflection on the rising branch of their light curves, just before
maximum light.
Furthermore, the  $\omega$
Cen stars
for which K97 found periods of about 1 day are all more than $0.5$ mag brighter
than the RR Lyrae variables. Since V10 is only $0.2$ magnitudes brighter than 
the mean $<V>$ for the other variables, we assume that our adopted period is
the correct one.

\subsection{Period-Luminosity and Period-Amplitude Relations}

The period-luminosity and period-amplitude relations are plotted in Figure 3.
In the upper panel of the diagram, the stars separate into three
separate regions which we interprete to be due to different modes of
pulsation: fundamental, first and second-overtone. W90 has already noted that
the variables with periods $\sim$0\day 34 may be pulsating in the 
second-overtone mode. Another interesting feature of Figure 3 is its 
similarity to the P-L and P-A plots that Clement \& Rowe (2000, hereafter CR)
made for the $\omega$ Centauri variables observed by K97. 
To illustrate this, we plot in Figure 4
the Fourier phase differences $\phi_{21}$, $\phi_{31}$ and the $V$
amplitude against $\log P$ for the NGC 5897 variables 
and include the $\omega$ 
Cen variables that CR considered to belong to the Oosterhoff (1939, 1944)
type II class 
(i.e. $<V>\le 14.65$ for the fundamental mode and $<V>\le 14.60$ for the 
overtone modes).
Also included is star 96 ($\log P= -0.3$, $A_V=0.17$) of K97 which was not 
included in CR's study. 

\begin{figure}
\epsscale{0.3}
\plotone{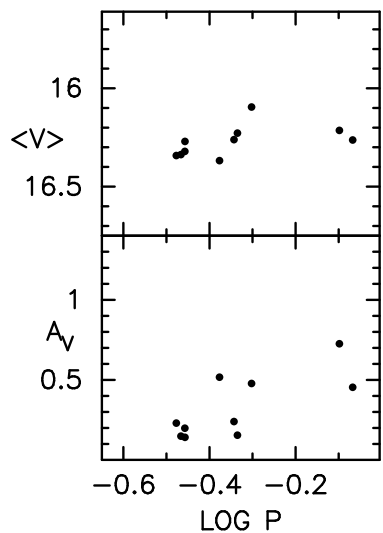}
\caption[]
{The period-luminosity and  period-amplitude 
relation ($<V>$ and $A_V$ versus $\log P$) for the RR Lyrae variables in 
NGC 5897.
\label {Fig. 3}}
\end{figure}

In Figure 4, it can be readily seen that the RRab stars V1 and
V6 have periods, Fourier phase differences
and amplitudes comparable to some of the $\omega$ Cen 
Oosterhoff type II fundamental mode pulsators. What is striking though
is that V1 and V6 have longer periods and hence
lower amplitudes than most of the $\omega$ Cen stars.
A similar trend can be seen among the first-overtone variables.
Both V2 and V11, with periods $\sim$0\day 46 ($\log  P \sim -0.34$), 
have much lower
amplitudes than most of the $\omega$ Cen first-overtone pulsators. The
case of V2 is particularly interesting because W90 found an increase in
period and a decrease in amplitude for this star
between the 1950s and the 1980s.  The $V$
amplitude we have derived for V2 ($0.24$ mag) is comparable to her
lower, more recent value ($0.22$ mag), but  the $V$ amplitude that 
she derived for the 1956-1966 observations was $0.39$ mag,
similar to the amplitudes of the $\omega$ Cen first-overtone
variables. V2 seems to have
changed significantly during the last 50 years. We analysed our observations
to search for evidence that V2 is in the process
of mode switching. To do this, we measured the 
residuals to the 
light curve and searched for periods in the range between 0\day 25
and 0\day 65. No oscillations with an amplitude greater than $0.03$ mag
were detected for any period in the interval. If the star were in the 
process of changing from first-overtone
to fundamental mode pulsation, we would have expected to detect oscillations
with a period $\sim$0\day 60. We therefore conclude that if V2 is in the process
of switching modes, the fundamental mode oscillations are still very weak.

\begin{figure}
\epsscale{0.3}
\plotone{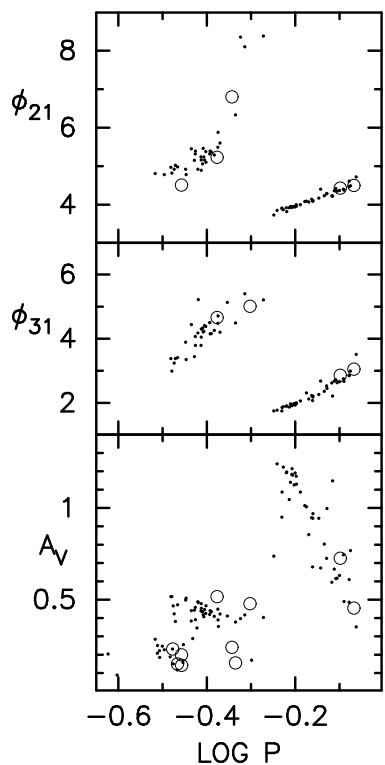}
\caption[]
{Plots of $\phi_{21}$, $\phi_{31}$ and
$A_V$ versus $\log P$ for the RR Lyrae variables in 
NGC 5897 (open circles) and the Oosterhoff type II RR Lyrae variables in
$\omega$ Centauri (dots). Values for $\phi_{21}$ and $\phi_{31}$ are 
plotted only for stars for which the error is less than $0.2$.
\label {Fig. 4}}
\end{figure}

The four stars with P$\sim$0\day 34 (V7, V8, V12 and V13) have periods
and amplitudes comparable to some of the $\omega$ Cen second-overtone variables,
but, like the RRab stars, they seem to congregate at the long period
end of the sequence in the P-A plot.
CR found that several of the second-overtone variables in $\omega$ Cen
exhibited non-radial pulsations. We performed a period search on the 
residuals to the
light curves of the above-mentioned four stars in an attempt to search for 
non-radial 
pulsations. No oscillations with amplitudes greater than $0.03$ mag were
detected in any of them. However,
the amplitude of the non-radial pulsation detected
by CR for star 186 in $\omega$ Cen was $0.10$ mag, much higher than this.

As noted previously, the 
light curve for V10 has a structure similar to the light curves
of V47, V68 and V123 in $\omega$ Cen and we suggest that 
it might be an anomalous Cepheid. The properties of anomalous Cepheids (ACs)
have been discussed by Nemec et al. (1994, hereafter N94). 
They are located in the Cepheid instability strip and
are more luminous than horizontal branch stars, but they tend to have 
shorter periods than globular cluster Cepheids
because they are more massive. 
They are probably
coalesced binary stars and may be related to blue stragglers.
Most of the known ACs occur in nearby dwarf galaxies, but N94
pointed out that V68 in $\omega$ Cen might be an AC pulsating in the
fundamental mode.\footnote{N94 also
considered that V84 in $\omega$ Cen
might be an AC candidate, but in
the meantime, van Leeuwen et al. (2000) have shown that
V84 is not a member of $\omega$ Cen.} 
We assume that if V68 is an AC, then V47 and V123 in $\omega$ Cen
should also be classified as ACs and if 
clusters with blue stragglers
are more likely to contain ACs, then it is plausible that there are ACs
in NGC 5897 as well. 
Testa et al. (2001) pointed out that the cluster has a sparse, but clearly
visible population of blue stragglers. If V10 is an AC, then V3 could be one
as well.
N94 derived a slope for the period luminosity
relation for AC variables: $\Delta V/\Delta \log P=-3.13\pm 0.28$ and
if both V3 and V10 are ACs, V3 should be $0.23$ mag
fainter than V10. The actual difference is $0.27$ mag, in reasonable agreement
with the prediction of N94. Another possibility raised by N94 is that
stars like this might be first-overtone population II Cepheids. Thus, there
is still some doubt about the true nature of V3 and V10.

Our study reinforces the conclusion
that the period distribution of the RR Lyrae variables
in NGC 5897 is  unusual. W90 plotted a period-frequency distribution
that demonstrated that the cluster has no
variables with periods between 0\day 45 and 0\day 75 where the
majority of RRab variables are found in other clusters. Although we
have discovered
four new variables in this investigation, and two of
them have periods between 0\day 45 and 0\day 50, neither is an RRab star.
Furthermore,
there are still no known variables with periods between 0\day 50 and 0\day 75
and most cluster RRab stars have periods in this range.

%
%
\acknowledgements

We thank Adam Muzzin for his assistance in the preparation of the paper.
Support for this work from
the Natural Sciences and Engineering Research Council of Canada is
gratefully acknowledged.

%

\begin{deluxetable}{ccccc}
\tablecaption{ A Sample of the Observations Obtained on May 1\label{Table 1} }
\tablewidth{0pt}
\tablehead{
\colhead{Frame} & \colhead{Frame} & \colhead{Frame} & \colhead{Combined} 
&\colhead{mean} 
\\
\colhead{\#1} & \colhead{\#2} & \colhead{\#3} & \colhead{Frame \#} 
&\colhead{Hel. JD} }
\startdata
 971095 & 971096 & 971097 & may01add01 & 2450570.5335 \nl
 971096 & 971097 & 971098 & may01add02 & 2450570.5373 \nl
 971097 & 971098 & 971099 & may01add03 & 2450570.5410 \nl
 971098 & 971099 & 971100 & may01add04 & 2450570.5439  \nl
 971099 & 971100 & 971101 & may01add05 & 2450570.5466  \nl
 971100 & 971101 & 971102 & may01add06 & 2450570.5495 \nl
\enddata
\end{deluxetable}

\begin{deluxetable}{cccccclcccc}
\tablecaption{ The RR Lyrae Variables in NGC 5897 \label{Table 1} }
\tablewidth{0pt}
\tablehead{
\colhead{Var} & \colhead{SK}& \colhead{$x$} & \colhead{$y$} 
& \colhead{$x"$} & \colhead{$y"$} & \colhead{Period}  
& \colhead{$<V>$} & \colhead{$A_V (\sigma)$} & $\phi_{21} (\sigma)$  
& $\phi_{21} (\sigma)$ \\
\colhead{\#} & \colhead{\#}& \colhead{(pix)} & \colhead{(pix)} 
& \colhead{} & \colhead{}& \colhead{(days)}   
& \colhead{} &\colhead{} &\colhead{} &\colhead{}   
}
\startdata
 1 & 351 & 125 & 688 & -109 & -201 & 0.797296 & 16.21 & 0.73 (0.034)
 & 4.43 (0.04) & 2.86 (0.05) \nl
 2 & 299 & 347 & 498 &  -57 &  -97 & 0.453945 & 16.26 & 0.24 (0.024)
 & 6.80 (0.17) & 5.97 (0.49) \nl
 3 & 206 & 556 & 397 &  -40 &   -4 & 0.419917 & 16.37 & 0.52 (0.026)
 & 5.23 (0.19) & 4.66 (0.17) \nl
 4 &     & 549 & 116 &  +71 &  +20 & 0.83127  &  ---  & ---   &  ---        & 
  --- \nl
 6 & 118 & 673 & 221 &  +16 &  +59 & 0.856223 & 16.26 & 0.45 (0.026)
 & 4.50 (0.04) & 3.05 (0.08) \nl
 7 & 161 & 608 & 203 &  +31 &  +35 & 0.348931 & 16.32 & 0.20 (0.024)
 & 4.51 (0.18) & 2.73 (0.34) \nl
 8 &  63 & 800 &  80 &  +58 & +122 & 0.341802 & 16.34 & 0.15 (0.025)
 & 4.59 (0.51) & 6.78 (0.56) \nl
 10 & 177 & 609 & 444 & -64 &  +12 & 0.49837  & 16.10 & 0.48 (0.022)
 & 4.45 (0.26) & 5.01 (0.09) \nl
 11 & 120 & 649 & 219 & +20 &  +49 & 0.46233  & 16.23 & 0.16 (0.025)
 & 2.60 (0.21) & 7.70 (0.61) \nl
 12 &     & 127 & 235 & +70 & -158 & 0.3333   & 16.34 & 0.23 (0.024)
 & 5.38 (0.41) & 4.49 (0.75) \nl
 13 & 116 & 742 & 238 &  +2 &  +84  & 0.3491   & 16.27 & 0.14 (0.029)
 & 1.61 (1.17) & 6.71 (0.63) \nl
\enddata
\end{deluxetable}

\begin{deluxetable}{cc}
\tablecaption{ Observations for V1 \label{Table 3}}
\tablewidth{0pt}
\tablehead{
\colhead{Mean Hel. JD} & \colhead{$V$} \\
\colhead{-2,450,500} & \colhead{} 
}
\startdata
 70.665 & 16.410 \nl
 70.673 & 16.405 \nl
 70.680 & 16.394 \nl
 70.685 & 16.415 \nl
 70.779 & 16.491 \nl
 71.548 & 16.459 \nl
 71.564 & 16.439 \nl
 71.568 & 16.442 \nl
\enddata
\end{deluxetable}

\end{document}